# An Analytical Model of Wavelength-dependent Opposition Surge in Emittance and Reflectance Spectroscopy of Airless Rocky Exoplanets

Leonardos Gkouvelis
Ludwig Maximilian University, Faculty of Physics, University Observatory, Scheinerstrasse 1, Munich D-81679, Germany; leo.gkouvelis@physik.lmu.de

## Abstract

Theoretical frameworks for reflection and emission spectroscopy of exoplanet surfaces are becoming increasingly important for the characterization of rocky exoplanets, especially with the rapid growth of the detected exoplanet population and observational capabilities. The Hapke theory of reflectance and emittance spectroscopy has been widely adopted in the exoplanet community, yet a key physical effect—the opposition surge enhancement at small phase angles—remains largely neglected. This phenomenon, driven by shadow hiding and coherent backscattering, introduces a significant brightening that depends on wavelength, particle size, and surface morphology. In this paper, I propose an alternative formulation for opposition surge modeling, ensuring a smooth-to-sharp transition at small phase angles, dictated by wavelength-dependent scattering properties. I evaluate the impact of opposition surge on phase curves and surface spectra, comparing a family of models with increasing simplifications, ranging from a full wavelength-dependent opposition effect to its complete omission. My results indicate that neglecting opposition effects can introduce systematic deviations in retrieved albedos, spectral features, and phase curves, with errors reaching up to 20%–30% in certain spectral bands. Upcoming JWST observations will probe phase angles below ∼10° for rocky exoplanets around M dwarfs; thus, accounting for opposition effects is crucial for accurate surface characterization. Proper treatment of this effect will lead to improved retrievals of surface albedo, mineralogical composition, and roughness properties. This study establishes a physically consistent framework for exoplanet phase-curve modeling and provides a foundation for future retrieval algorithms aimed at interpreting exoplanet surfaces.

*Unified Astronomy Thesaurus concepts:* Radiative transfer (1335); Extrasolar rocky planets (511); Spectroscopy (1558)

## 1. Introduction

The population of detected exoplanets has grown significantly in recent decades; however, despite the rapid evolution of the field, the fraction of planets characterized as rocky worlds corresponds to a tiny fraction considering objects with radii $<2\,R_\oplus$ and masses $<10\,M_\oplus$, including Earth-like exoplanets and high-density super-Earths (C. D. Dressing & D. Charbonneau 2015; B. J. Fulton et al. 2017; T. A. Berger et al. 2018). To date there has not been a confirmed detection of an atmosphere for a rocky exoplanet although there have been a few attempts (e.g., L. Kreidberg et al. 2019) with the Spitzer InfraRed Array Camera (IRAC) and, more recently, with the James Webb Space Telescope (JWST; see, e.g., T. P. Greene et al. 2023; S. Zieba et al. 2023; Q. Xue et al. 2024; A. Bello-Arufe et al. 2025) observing the thermal emission of the 15 $\mu$m CO$_2$ absorption band and 12.8 $\mu$m band (E. Ducrot et al. 2024) with the Mid-Infrared Instrument (MIRI), and other promising candidates are accepted for observations (e.g., B. Benneke et al. 2024). In the case of no atmosphere or a very thin atmosphere (less than ≈0.01 bars) and hence no significant heat redistribution across the planet, we can consider the so-called "bare-rock" scenario, where the observation comes directly from the surface. The interpretation of future observations and the characterization of those rocky exoplanets are subject to the forward models we are applying to invert the problem and retrieve the physical parameters of the atmosphere or surface of the exoplanet. In the case of bare rock, the spectrum is produced from reflected light and thermal emission from the surface. For close-in tidally locked exoplanets, which are easier to observe, the high surface temperatures cause reflected light and thermal emission to overlap in the spectral range, necessitating a model that combines both contributions. With the current and near-future observational capabilities of telescopes like JWST and extremely large telescopes (ELTs), there is a growing need to develop robust models for interpreting the reflected light and thermal emission spectra from upcoming observations (N. E. Batalha et al. 2019; B. Benneke et al. 2019). The study of exoplanet phase curves has played a fundamental role in characterizing planetary atmospheres and surfaces. Unlike solar system observations, where spatially resolved photometry is possible, exoplanet studies rely on disk-integrated light curves that vary with the planet's orbital phase. The analysis of these phase-dependent brightness variations has provided constraints on planetary albedo, atmospheric circulation, and cloud properties (V. Parmentier & I. J. M. Crossfield 2018). With the advent of the Kepler mission, phase curves of numerous exoplanets were obtained, allowing statistical studies of their optical properties (L. J. Esteves et al. 2013, 2015). These works demonstrated that planetary brightness varies significantly across different systems, with some planets exhibiting high geometric albedos due to clouds or hazes, while others appear darker, likely due to strong molecular absorption. Subsequent observations with Hubble Space Telescope (HST) and JWST have provided spectrally resolved phase curves, improving our understanding of thermal emission and reflected light (L. Kreidberg et al. 2019; S. Zieba et al. 2022). These studies have been instrumental in identifying temperature variations and cloud dynamics in hot Jupiters such as WASP-121b and KELT-1b (M. Lendl et al. 2020; G. Scandariato et al. 2022). More recently, high-precision optical photometry from the CHEOPS mission has extended phase-curve measurements to new planetary classes, including ultrahot and highly irradiated exoplanets (M. Lendl et al. 2020). Despite these advancements, the range of phase angles typically observed remains limited. Most exoplanet phase curves are constrained to phase angles $\alpha \gtrsim 30°$, due to observational limitations in transit and secondary eclipse methods. Only in rare cases, such as those observed by JWST, do data sets extend to small phase angles ($\alpha \lesssim 10°$), where opposition effects become significant (S. Zieba et al. 2022). This highlights a key challenge in exoplanet photometry: while existing models often assume isotropic scattering or simple phase functions, real planetary surfaces and atmospheres exhibit complex angular-dependent reflectance properties. Theoretical modeling of exoplanet phase curves has employed a variety of approaches beyond the Hapke model, including Lambertian reflection (S. Seager et al. 2000) and Chandrasekhar's radiative transfer solutions (S. Chandrasekhar 1960). In exoplanet science, the construction of models differs from the solar system approach and from the interpretation of remote-sensing observations. The star–planet system is a point source, so that (1) we can solely observe the star during secondary eclipse and (2) the reflected or emitted radiation from the planet reaches the observer from all visible longitudes and latitudes of the planet. This means that reflection must be treated at various angles, as each surface element has a different temperature, receives and reflects radiation differently, and thus emits differently. To model a surface spectrum, we first need to consider the surface material composition. At sufficiently high temperatures, around ≈1600 K, the planetary surface undergoes melting, and its thermal emission increasingly resembles that of a blackbody. Observations of active lava flows on Earth, including in situ measurements in Hawaii and laboratory experiments, support this behavior. These studies indicate that near-infrared spectra of molten lava are largely dominated by blackbody radiation (e.g., L. P. Flynn & P. J. Mouginis-Mark 1992; V. Lombardo et al. 2020). At lower temperatures, the crust interacts with electromagnetic radiation through a combination of reflection, absorption, scattering, and emission, which depends on the geometry of the interaction, the physical character of the surface, the wavelength of radiation, and the microcrystalline nature of the material.

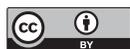

Extracting the optical properties of a surface, such as the single-scattering albedo (which represents the fraction of scattered light relative to total extinction; B. Hapke 1993), is inherently challenging from disk-integrated observations alone, even with wide phase-angle coverage (D. L. Domingue et al. 1991). Optical properties, including the single-scattering albedo ($\omega$), refractive index ($n, k$), and phase function, describe the fundamental light-scattering behavior of surface materials and determine how light interacts at the microscopic level. These properties govern the bidirectional reflectance distribution function (BRDF), which characterizes reflectance as a function of incident and viewing angles. In contrast, photometric parameters describe the observed brightness of an exoplanet at different phase angles, integrating optical properties over the entire planetary disk. These include the geometric albedo ($A_g$), which quantifies reflectance at zero phase angle, and the Bond albedo ($A_B$), which represents the total fraction of scattered light. The phase curve, which describes brightness variations with phase angle, encodes both optical properties and macroscopic effects such as surface roughness and multiple scattering. In this study, I introduce a modeling framework that explicitly accounts for opposition surge effects in reflectance and emission spectra of airless exoplanets. By incorporating optical properties from laboratory measurements and evaluating their impact on photometric parameters, we ensure a physically consistent approach to interpreting exoplanet observations. Since exoplanet observations are inherently limited in quality (and will likely remain so for the foreseeable future), we must work inversely by making informed assumptions about surface materials and utilizing known optical properties. While ab initio calculations do not exist for every mineral phase, numerous laboratory databases provide measured reflectance spectra at various angles (R. Milliken 2020, R. N. Clark et al. 2007). The Hapke theory of reflectance and emittance spectroscopy is an analytic model that provides a useful framework to evaluate the optical properties of individual surface materials, such as minerals, regolith grains, and their mixtures, and their potential effects on spectral features. These optical properties include the single-scattering albedo ($\omega$), phase function, opposition effect parameters, and macroscopic roughness, all of which influence the observed reflectance and emittance spectra (B. Hapke 1981, 1984, 1986, 2002; B. W. Hapke et al. 1993; B. Hapke et al. 1998; B. Hapke 2005, 2008, 2021). Using the Hapke model, I retrieve these optical properties by applying it to bidirectional reflectance laboratory measurements from spectral databases (e.g., R. Milliken 2020; R. N. Clark et al. 2007). Once retrieved, I reapply them within the same framework to simulate exoplanet surfaces and assess their spectral and photometric properties. The opposition surge is a well-known photometric phenomenon observed at small phase angles, characterized by a rapid increase in reflectance as illumination and observation directions align. A classic astronomical example is the lunar surface, where this effect has been extensively studied and explained by two primary physical mechanisms: shadow hiding (SHOE) and coherent backscattering (CBOE; B. Hapke 2012). Both mechanisms depend on microphysical surface properties, including particle size, composition, porosity, and wavelength of observation (B. Hapke 1993; R. M. Nelson et al. 2000; Y. G. Shkuratov & P. Helfenstein 2001). Notably, the lunar surface observations demonstrate variations in the angular shape and amplitude of this phenomenon across different regions. For instance, Apollo-era observations and later photometric studies revealed significant differences in opposition surge characteristics between lunar highlands, mare regions, and impact ejecta zones, reflecting variations in particle size distributions, porosity, and microscopic surface textures (see, e.g., B. J. Buratti et al. 1996; P. Helfenstein & M. K. Shepard 1999; Y. Shkuratov & L. Starukhina 1999; V. Kaydash et al. 2013). Similarly, observations of other airless solar system bodies, such as Mercury, asteroids, and icy satellites, have consistently shown opposition surges that vary depending on local surface properties and particle microstructure (D. Domingue & A. Verbiscer 1997; R. M. Nelson et al. 1998; B. Hapke 2012). Thus, the opposition surge provides a diagnostic window into the microphysical properties of planetary surfaces.

In this study, I develop a physically motivated modeling framework designed specifically for rocky exoplanets that explicitly incorporates the opposition surge effect using laboratory-derived optical and photometric properties. This enables more realistic simulations of reflectance and emission spectra, improving the interpretation of upcoming rocky exoplanet observations. To adequately represent the complexity of the opposition effect observed in solar system surfaces, I introduce a flexible mathematical treatment of the opposition surge that smoothly transitions between Gaussian and Lorentzian profiles, guided exclusively by the laboratory-derived optical data. This approach allows for wavelength-dependent characterization of the opposition surge contribution to the secondary eclipse depth, and it is especially well suited to future Bayesian retrieval frameworks aimed at interpreting rocky exoplanet observations.

## 2. The Opposition Surge Effect

The opposition surge effect was introduced to fit observational data at small phase angles that show a sudden brightening (B. Hapke et al. 1996; B. Hapke & H. van Horn 1963; W. W. Montgomery & R. H. Kohl 1980; R. M. Nelson et al. 1998, 2000; R. G. French et al. 2007). Small-scale roughness or porosity on surfaces induces the opposition effect, which arises from a combination of two distinct physical mechanisms: SHOE and CBOE. SHOE occurs as a result of the progressive elimination of cast shadows as the phase angle approaches zero. This effect is primarily governed by the geometric arrangement of grains and their packing density, and in its simplest form, it is independent of wavelength. However, if the grain size is comparable to or larger than the wavelength, scattering properties such as diffraction and absorption introduce an indirect wavelength dependence. In contrast, CBOE is a wave interference phenomenon that occurs at the photon level after multiple scattering events within the medium. The width of the CBOE surge is strongly dependent on wavelength, as constructive interference enhances backscattering most effectively when the path length differences between scattered photons approach integer multiples of the wavelength (B. Hapke 1993; B. Hapke et al. 1998; B. Hapke 2012; K. Muinonen et al. 2002). Experiments have shown that the effect is pronounced for fine powders with particles with mean size $\lessapprox 20\,\mu m$ (B. Hapke 2012) and mean albedo $\sim 0.4$ (R. M. Nelson et al. 2004). The removal of shadows in SHOE is governed not by interference but by geometric constraints that lead to a gradual descent similar to a Lorentzian or hyperbolic function. The size and structure of the grain particles may introduce wavelength dependence in the sense that scattering properties of the particles change with wavelength. Larger particles produce larger shadows, and if the wavelength is comparable to the particle size, it can enhance the effect.

CBOE effect is strongly dependent on wavelength because it arises from the interference of light waves scattered by particles. The constructive interference arises from photons that are scattered along nearly identical but opposite directions. As the phase angle approaches zero, the optical path lengths of scattered photons become nearly identical, leading to constructive interference that enhances the intensity of the scattered light. Both phenomena are combined to explain observations in solar system bodies (Z. M. Dlugach & M. I. Mishchenko 2013; B. W. Hapke et al. 1993; B. Hapke et al. 1998; K. Muinonen 2002; K. Muinonen et al. 2002). The exact peak shape and angular width of the opposition surge are shifting between Gaussian and Lorentzian profiles according to scattering dynamics of the wavelength and material properties (R. M. Nelson et al. 1998; B. Hapke et al. 1998; I. G. Shkuratov 1988; Y. G. Shkuratov & P. Helfenstein 2001). The shape of the opposition surge varies between Gaussian and Lorentzian profiles, depending on the scattering dynamics, wavelength, and material properties. In the case of CBOE, the available scattering paths within a medium of varying thickness influence the degree of constructive interference, resulting in either a sharp, narrow peak or a broader, less intense one (E. Akkermans et al. 1988). For SHOE, surfaces composed of a mixture of particle sizes tend to produce a broad, symmetrical peak that is well described by a Gaussian profile. In contrast, sparsely arranged large particles create a steeper phase curve, more characteristic of a Lorentzian shape (D. G. Stankevich & Y. G. Shkuratov 2000; Y. G. Shkuratov et al. 2005). Evidently, the shape of the opposition peak can provide information about the depth and density of the surface or any particulate medium. For example, B. J. Buratti et al. (1996) demonstrate that observations from different geographical locations on the Moon show variations in the surge peak shape. Several independent observations indicate that the zero-phase peak is caused by both SHOE and CBOE in roughly equal amounts. The relative contribution of each effect depends on surface properties such as particle size, porosity, composition, and the wavelength of the radiation.

In addition to SHOE and CBOE, a third phenomenon is producing a systematic increase in reflectance due to macroscopic (large-scale) roughness features. By macroscopic roughness, I refer to surface roughness with a length scale that far exceeds the particle size of the medium but remains unresolved by the imaging system. Experimental studies by A. M. Cord et al. (2003, 2005), observations of the lunar surface by P. Helfenstein & M. K. Shepard (1999), and investigations of terrestrial volcanic terrains by S. Labarre et al. (2015, 2017, 2019) indicate that the most photometrically influential roughness scales can vary from submillimeter- to centimeter-sized features. The macroscopic roughness backscattering bias (MRBB) is a key challenge to accurately model the influence of surface roughness on bidirectional reflectance. As noted by M. K. Shepard & B. A. Campbell (1998), parameterizing surface roughness is complex because surface statistical properties, such as rms heights and slopes, depend on the measurement scale. Various methods have been developed to account for the photometric effects of macroscopic roughness on planetary surfaces. Some models utilize geometrically defined structures, such as crater-like depressions or cavities (J. van Diggelen 1959; J. Veverka & L. Wasserman 1972; K. Lumme & E. Bowell 1981a, 1981b), while others rely on statistical descriptors or fractal-based approaches (B. Hapke 1984; D. Despan et al. 1998, 1999; M. K. Shepard & B. A. Campbell 1998). The most widely used model for the photometric effect of MRBB of planetary bodies was introduced in B. Hapke (1984), and an alternative was proposed recently by D. J. Shiltz & C. M. Bachmann (2023), where they provide a more accurate model for phase angles less than 90°. MRBB is distinct from the opposition surge contributions because it introduces a bias across all phase angles but is also enhanced at small phase angles. Since my current work focuses on opposition effects rather than macroscopic roughness, I do not explicitly model MRBB. However, given its potential impact on exoplanet surface reflectance, future studies should consider incorporating a separate treatment of MRBB alongside SHOE and CBOE. The equation of radiative transfer does not treat the opposition surge, and it has to be introduced ad hoc (M. S. Bobrov 1962; B. Hapke 2012). This treatment has been the common practice for modeling and data analysis fitting. An analytical derivation for CBOE is given by E. Akkermans et al. (1986), and one for SHOE can be found in B. Hapke (2012). Both derivations lead to a Lorentzian shape; nevertheless, for the reasons mentioned above, in practice various types of profiles, changing from smooth to sharp, are used to fit observational data. Those two types of shapes can be fitted with Lorentzian or Gaussian mathematical functions (E. Akkermans et al. 1988; D. G. Stankevich & Y. G. Shkuratov 2000; B. Hapke 2012). Furthermore, the shape of the opposition surge can change across wavelength. It can change due to each of the two phenomena individually, but also as a result of their combined effect, since their relative dominance varies across the spectrum. In the description of the complete Hapke theory, the contribution of each phenomenon is additively combined in the bidirectional reflectance equation, where it is introduced ad hoc (B. Hapke 2012). This approach forces the user to choose a priori a mathematical function for each of the two phenomena, without having real knowledge on the actual shape of the applying material. This is even worse when we introduce the wavelength dependence of the surge shape. In theory, one could choose the proper shape function by fitting to laboratory data with high-resolution multiangular measurements for angles approaching zero. In practice, this is difficult to perform for hundreds or even thousands of combining materials and nearly impossible if we introduce multiwavelength coverage. In exoplanet science, where we are agnostic on the composition and the terrain microstructure (e.g., spacing of larger grains), we want a model that handles both phenomena in a flexible manner as a function of wavelength. In this work, I introduce an ad hoc mathematical tool—described later in the manuscript—that does not impose a predefined shape on the two phenomena. Instead, it is capable of adopting profiles ranging from smooth to sharp, with wavelength dependence determined by the optical properties retrieved from laboratory measurements of each material. The shape of the opposition surge is then driven exclusively by the laboratory material data as a function of wavelength. I demonstrate that this approach has a smooth numerical behavior and is optimal for being used as a forward model in future inversion techniques and data fitting algorithms. Furthermore, as these types of theoretical frameworks—namely, various approximations of the Hapke model that often neglect the opposition effect—are gaining popularity in exoplanet science for simulating rocky surfaces, I compare a family of models with increasing complexity and highlight their differences based on the underlying assumptions. I discuss the advantages and limitations of each approximation and propose potential future applications. Finally, I emphasize the important role that such frameworks can play as part of a Bayesian analysis pipeline for the characterization of rocky exoplanets.

## 3. Theory

Let me first define the background formulation in order to construct a family of models, and the reader can find their derivations in their references accordingly. The exoplanet–star system, typically, is observed as a point source. Reflected light and thermal emission might overlap on hot planets, so we have to apply a unified treatment. To estimate the emitted light of the whole disk, we can follow the reasoning from Y. G. Shkuratov & P. Helfenstein (2001) and the adaptation for exoplanets by the work of R. Hu et al. (2012). I can define a mesh-grid where each surface unity has a set of angles, namely, incident angle $i$, emergent angle $e$, and phase angle $g$, such that the total thermal contrast will be the integral across all visible longitude–latitude surface elements:

$$\frac{F_p}{F_*} = \frac{1}{F_{\rm inc}} \int_{-\frac{\pi}{2}}^{\frac{\pi}{2}} \int_{-\frac{\pi}{2}}^{\frac{\pi}{2}} I_p(\theta, \phi) \cos^2\theta \cos\phi \, d\theta d\phi \times \left(\frac{R_p}{D_p}\right)^2. \quad (1)$$

I set the geometric transformation relations between the star–planet system and the system–observer system to be

$$\mu_0 \equiv \cos i = \cos\theta \cos(\alpha - \phi) \quad (2)$$

$$\mu \equiv \cos e = \cos\theta \cos\phi \quad (3)$$

$$g = \alpha, \quad (4)$$

where I define the orbital phase angle ($\alpha$) such that $\alpha = 0°$ corresponds to opposition, when the planet is behind the star during secondary eclipse, and $\alpha = \pm 180°$ corresponds to transit. The stellar phase angle ($g$) is the angle between the observer, the planet, and the host star. By default the stellar coplanar radiation comes from the direction of ($\theta = 0, \phi = \alpha$). The incident radiation at each surface element is

$$F_{\rm inc} = \pi B_\lambda[T_*]\left(\frac{R_*}{D_p}\right)^2, \quad (5)$$

where $B$ is the wavelength-dependent stellar Planck function, $R_*$ is the stellar radius, and $D_p$ is the star–planet distance. The total incoming radiation from each surface element is a combination of reflected, scattered, and thermally emitted radiation and expressed in Equation (1) as the $I_p$ function:

$$I_p(\theta, \phi) = I_s(\theta, \phi) + I_t(\theta, \phi), \quad (6)$$

where

$$I_s(\theta, \phi) = \frac{F_{\rm inc}\mu_0}{\pi} r_c(\mu_0, \mu, g) \quad (7)$$

$$I_t(\theta, \phi) = \epsilon(\mu) B_\lambda[T(\theta, \phi)]. \quad (8)$$

In Equations (7) and (8), two important functions are introduced. The first parameter, $r_c$, represents the reflectance factor, which quantifies the brightness of a surface relative to a Lambertian surface under identical illumination conditions (M. K. Shepard et al. 2017). A Lambertian surface is an idealized surface that scatters incident light isotropically in all directions, following Lambert's cosine law, where the observed intensity is proportional to the cosine of the incident angle (S. Chandrasekhar 1960; S. Seager & J. J. Lissauer 2010). For a Lambertian sphere, $r_c = 1$, while real planetary surfaces exhibit anisotropic scattering due to microscopic and macroscopic roughness effects. While the BRDF itself is governed by surface geometry and does not directly depend on chemical composition, the optical properties of a material, such as its refractive index and absorption coefficient, can influence the overall reflectance behavior. These material-dependent properties, combined with surface roughness, determine deviations from purely Lambertian scattering. The second function is $\epsilon$, the emissivity, which characterizes how a surface with temperature $T_s$ deviates from behaving as an ideal blackbody. Specifically, it defines how the material emits radiation compared to the Planck function $B_\lambda[T_s]$.

B. Hapke (1981, 2002) provides an analytical approximation for the radiance coefficient of any particulate material in terms of its single-scattering albedo, $\omega$, which quantifies the fraction of radiation scattered or reflected in a single-scattering event and is defined as $\omega = \frac{\sigma_{\text{scat}}}{\sigma + \sigma_{\text{scat}}}$, where $\sigma$ and $\sigma_{\text{scat}}$ are the absorption and scattering cross-sections, respectively:

$$r(i, e, g) = \frac{\omega}{4} \frac{\mu_0}{\mu_0 + \mu}[P(g) + H(\mu_0)H(\mu) - 1][1 + S]. \quad (9)$$

This solution assumes isotropic scatterers for multiple scattering and nonisotropic scatterers for single scattering (B. Hapke 1981). The function $P(g)$, known as the phase function, is expressed as an infinite sum of Legendre polynomials $P_n(\cos g)$:

$$p(g) = 1 + \sum_{n=1}^{\infty} b_n P_n(\cos g) \approx 1 + b \cos g. \quad (10)$$

Numerical experiments show that calculating the series and comparing it with the first-order Legendre polynomial expansion of $p(g)$ results in practically negligible differences for all the simulations presented in this work. The functions $H(x)$ are the Ambartsumian–Chandrasekhar H-functions (V. A. Ambartsumyan 1958; S. Chandrasekhar 1960), which approximate multiple scattering effects and are given by

$$H(x) = \frac{1 + 2x}{1 + 2x\sqrt{1 - \omega}}. \quad (11)$$

In Equation (9), $i$ and $e$ represent the incidence and emission angles, respectively, while $g$ is the phase angle, which is the angular separation between the incident and scattered light directions. The function $P(g)$ describes the phase function, which governs the angular distribution of scattered light. The terms $H(x)$ correspond to the Ambartsumian–Chandrasekhar H-functions, which provide an approximation for multiple scattering effects within the medium. Finally, $S$ represents the opposition surge enhancement term, which accounts for the characteristic brightening observed at small phase angles. $S$ is the total additive effect of two contributions as $S = S_S + S_C$ for SHOE and CBOE accordingly. It is fundamentally connected to the microscopic-scale structure of the surface, such as individual grains, crystals, or particles within a regolith or particulate surface layer. It is not directly connected to macroscopic structures like stones, rocks, boulders, or mountains. These particles, which can be as small as microns to millimeters, are often irregular in shape and form a layer of loose material with a porous structure. The opposition surge is created because these particles cast shadows on each other, and those shadows are hidden when the light source is directly behind the observer (P. Helfenstein 1988; Y. Shkuratov &

L. Starukhina 1999), or from coherent backscattered light as was already explained in the introduction section. The Hapke Henyey–Greenstein function (L. G. Henyey & J. L. Greenstein 1941) describes the opposition surge effect of SHOE as

$$S_S = B_0 \frac{1}{1 + \frac{1}{h}\tan\left(\frac{g}{2}\right)}, \quad (12)$$

where the subscript $S$ distinguishes the contribution to the opposition effect from SHOE. $B_0$ is the amplitude of the opposition effect, and $h$ is the angular width parameter that controls the spread of the opposition surge in phase-angle space. Mathematically this term is introducing a sharp shape around small phase angles to our phase curve. However, this function is not convenient for numerical calculations because the tangent function at small angles might lead to numerical instability and precision errors, while at large phase angles it grows rapidly, which makes it sensitive to numerical errors. A common practice is to approximate Equation (12) with a linear behavior at small phase angles or use alternative empirical functions that better capture the opposition surge effect, particularly when fitting observational data. As explained in the introduction, empirical evidence suggests that a Gaussian profile provides a more accurate fit to observed phase curves (B. Hapke 1986, 1993). This is because the sharpness and width of the opposition surge vary with surface properties, and a Gaussian formulation allows for a more flexible representation. While Equation (12) produces a sharper peak, the Gaussian formulation provides a smoother, more gradual transition, making it a better approximation for certain types of surfaces. Therefore, Equation (12) can be modified as

$$S_S = B_{0,S} e^{-\frac{g^2}{2h^2}}. \quad (13)$$

This substitution allows for a broader range of opposition surge shapes to be modeled, depending on the material properties and observational constraints. To account for CBOE, there is no rigorous theoretical derivation, but an approximate model for isotropic scatterers has been derived by E. Akkermans et al. (1986) and an empirical model has been derived by Y. G. Shkuratov & A. A. Ovcharenko (1998), which has a sharp peak that can be described as a hyperbolic or Lorentzian-type function (Equation (33) of B. Hapke 2002):

$$S_C = B_{0,C} \frac{1 + \frac{1 - e^{-\frac{1}{h_C}\tan(g/2)}}{(1/h_C)\tan(g/2)}}{2[1 + (1/h_C)\tan(g/2)]^2}, \quad (14)$$

with $h_C = \frac{\lambda}{4\pi\ell}$, where $\lambda$ is the wavelength and $\ell$ is the transport mean free path in the medium. However, if the photons are absorbed as they propagate between the scatterers, a property that depends on the single-scattering albedo $\omega$ of the material itself, few will be able to travel a long distance, so that the peak would be rounded off similar to a Gaussian profile (E. Akkermans et al. 1988). n theory, these two effects are independent of each other. In practice, however, both terms must be added to the surge term in Equation (9) to account for the contributions of each mechanism (B. Hapke 2012).

I continue by evaluating the emissivity as used in Equation (8). To do this, I first define the directional–hemispherical reflectance, which quantifies the fraction of incoming radiation that is scattered in all directions by the surface:

$$r_{\rm dh} = \frac{1 - \sqrt{(1-\omega)}}{1 + 2\sqrt{(1-\omega)}\mu_0}. \tag{15}$$

I also define the spherical reflectance, following B. Hapke (2005), as

$$r_s = \int_0^1 r_{\rm dh}(\mu_0)\mu_0\, d\mu_0 \simeq \frac{1-\sqrt{(1-\omega)}}{1+\sqrt{(1-\omega)}} \times \left[1 - \frac{1}{3}\frac{\sqrt{(1-\omega)}}{1+\sqrt{(1-\omega)}}\right]. \tag{16}$$

According to Kirchhoff's law, for a surface in thermal equilibrium, emissivity is directly related to the reflectance at a given wavelength. Thus, both Equations (15) and (16) can be used to estimate the emissivity of a surface element as

$$\epsilon_{\rm dir}(\mu) = 1 - r_{\rm dh}(\mu) \tag{17}$$

and

$$\epsilon_s(\mu) = 1 - r_s(\omega). \tag{18}$$

This assumption is valid for tidally locked rocky exoplanets without atmospheres, where thermal equilibrium holds across the surface. In order to close the system of the previous equations, I need to solve for the surface element temperature, and since Kirchhoff's law is obeyed, it follows that the energy balance can be written as

$$\int \mu_0 \epsilon_{\rm dir}(\mu_0) F_{\rm inc}\, d\lambda = \pi \int \epsilon_\lambda^h B_\lambda[T]\, d\lambda, \tag{19}$$

where $\epsilon_\lambda^h$ is the hemispherical average of the directional emissivity that I can evaluate as the double integral:

$$\epsilon_\lambda^h(\lambda) = \frac{1}{2\pi}\int_0^{2\pi}\int_0^1 \epsilon_{\rm dir}(\mu_0, \mu, \phi, \lambda)\, \mu\, d\mu\, d\phi, \tag{20}$$

with $\epsilon_{\rm dir}$ from Equation (15). Finally, I have to note that the integrals of Equation (19) are integrated accordingly where the downward and upward radiation is dominant, e.g., optical and infrared wavelengths for a solar system planet.

### 3.1. Wavelength Dependence Approach

I can ask the following question: How can one choose the shape of the opposition surge function a priori in theoretical models, particularly for rocky exoplanet surfaces? This is a critical issue because, in exoplanet science, surface material properties are typically assumed rather than directly measured. The challenge becomes even more significant when retrieving optical properties from laboratory measurements, which must be inverted using a model or directly fitted to data. In an ideal scenario, high-resolution multiangular measurements across wavelength space would allow for precise model selection. However, in practice, laboratory data sets often provide only a limited set of phase angles per material (R. E. Milliken et al. 2021; R. Milliken 2020). As a result, the opposition surge term in reflectance models can vary from negligible contributions to strong enhancements, depending on the material and observational conditions. In some cases it exhibits a smooth transition in phase curves, while in others it follows a sharp shape owing to stronger constructive interference effects. To address this variability, I propose a convolution-based approach to model the opposition surge. The Gaussian component accounts for smoother scattering effects, such as SHOE, while the Lorentzian component captures the sharp phase-curve enhancement due to CBOE. Their convolution ensures that the opposition effect is not arbitrarily forced into a specific shape but instead emerges dynamically as a function of wavelength, driven by the properties of the scattering medium. The convolution of a Gaussian function, $G(g')$, and a Lorentzian function, $L(g - g')$, provides a more flexible representation of the opposition surge. Mathematically, the total monochromatic opposition surge function is given by

$$S_V(g) = \int_{-\infty}^\infty G(g')L(g-g')\, dg', \tag{21}$$

where $g$ represents the phase angle, which is the angle between the incident and scattered light, and $g'$ is an integration variable used to describe the contributions from different scattering paths.

The Gaussian function $G(g')$ is

$$G(g') = \exp\left(-\frac{g'^2}{2h_G^2}\right), \tag{22}$$

where $h_G$ represents the width of the Gaussian component, controlling the smoothness of the opposition surge.

The Lorentzian function $L(g - g')$ is

$$L(g-g') = \frac{1}{1 + \left(\frac{g-g'}{h_L}\right)^2}, \tag{23}$$

where $h_L$ defines the width of the Lorentzian component, characterizing the sharpness of the opposition surge due to CBOE. To simplify the integral, I introduce the dimensionless variables:

$$u = \frac{g'}{\sqrt{2}\, h_G}, \quad \gamma = \frac{h_L}{\sqrt{2}\, h_G}. \tag{24}$$

These substitutions help express the convolution in a standard mathematical form:

$$S_V(g) = \frac{1}{\sqrt{2\pi}\, h_G}\int_{-\infty}^\infty \exp(-u^2)\frac{1}{1 + \left(\frac{g}{h_L} - \frac{\sqrt{2}\, h_G u}{h_L}\right)^2}\, du. \tag{25}$$

This integral is known in mathematical physics and can be related to the complex Faddeeva function (M. Abramowitz & I. A. Stegun 1964). To express the integral in terms of the Faddeeva function $w(z)$, I rewrite the term inside the Lorentzian function as

$$\frac{g}{h_L} - \frac{\sqrt{2}\, h_G u}{h_L} = \frac{g + i\gamma}{\sqrt{2}\, h_G}, \tag{26}$$

where I define the complex variable

$$z = \frac{g + i\gamma}{\sqrt{2}\, h_G}. \tag{27}$$

Thus, the final solution for $S_V(g)$ is given in terms of the real part of the Faddeeva function:

$$S_V(g) = \frac{B_0}{h_G\sqrt{2\pi}}\,{\rm Re}\,[w(z)], \tag{28}$$

**Table 1**
Construction of a Family of Models

| Name | Opposition | $\epsilon(\lambda)$ | $f(\lambda)$ |
|---|---|---|---|
| Sharp($\lambda$) | Equation (14) | Equation (17) | $\omega$, $B_0$, $h$, $b$ |
| Sharp | const. $B_0$, $h$, $b$ | Equation (17) | $\omega$, $B_0$, $h$, $b$ |
| Gaussian($\lambda$) | Equation (13) | Equation (17) | $\omega$ |
| Gaussian | const. $B_0$, $h$, $b$ | Equation (17) | $\omega$ |
| Convolved($\lambda$) | Equation (31) | Equation (17) | $\omega$, $B_0$, $h_L$, $h_G$, $b$ |
| Simple | ⋯ | Equation (17) | $\omega$ |
| Simple-$\epsilon_s$ | ⋯ | Equation (18) | $\omega$ |

where

$$w(z) = e^{-z^2} \text{erfc}(-iz) \quad (29)$$

and

$$z = \frac{g + i\gamma}{\sqrt{2}\, h_G}. \quad (30)$$

The final expression is

$$S_V(g) = \frac{B_0}{h_G \sqrt{2\pi}} \text{Re}\left[ w\left(\frac{g + i\gamma}{\sqrt{2}\, h_G}\right) \right], \quad (31)$$

where $B_0$ is the amplitude of the profile, $h_G$ is the Gaussian width, and $\gamma = \frac{h_L}{\sqrt{2} h_G}$. The Faddeeva function provides an efficient way to evaluate the opposition surge without direct numerical integration. While Equation (31) provides an exact analytical solution, it involves complex functions and rapidly oscillating integrals, making direct computation challenging. Instead, numerical evaluation methods are typically used for stability and accuracy. Thus, I modify Equation (9) by incorporating this wavelength-dependent opposition surge term, which allows for a flexible representation of the SHOE and CBOE contributions in the total reflectance. The effectiveness of this formulation will be tested in the following sections.

### 3.2. A Family of Models

Taking into account the complete theoretical framework described above, I define the solution of the closed system of equations by considering different approximation approaches for the bidirectional reflectance parameters. The parameters $\omega$, $b$, $h$, $h_L$, and $h_G$ are treated either as constants or as wavelength-dependent variables, depending on the model, with the latter two serving as amplitude factors in the newly formulated convoluted model. Each of these parameters is incorporated accordingly in the construction of a family of models, where emissivity is evaluated using either Equation (17) or Equation (18). I summarize the construction components of each approximation model in Table 1. As "Sharp($\lambda$)" I will define the model with sharp shape opposition surge and wavelength dependence on the optical parameters, and similarly for the "Gaussian($\lambda$)" and "Convolved($\lambda$)" models. In addition, I define the "Sharp" and "Gaussian" models with constant opposition surge term parameters $B_0$ and $h$. This approximation is similar to the work presented by R. Hu et al. (2012) and applied in L. Kreidberg et al. (2019) and S. Zieba et al. (2023) (The last two works apply for their synthetic spectra the model presented in R. Hu et al. 2012.) I discuss in the following section how I choose the values of the constants. In model "simple," I will take the assumption that the opposition effect is zero, an approximation that was followed by X. Lyu et al. (2024), and still keep emissivity the same as in previous models. For completeness, I will define an extra model, as "simple-$\epsilon_s$," with no surge term and emissivity to be evaluated from the spherical reflectance, $r_s(\lambda)$, as it is derived from Equation (18). This approximation was adopted by the recent work of M. Hammond et al. (2025). The optical properties of solid materials, like the single-scattering albedo, $\omega$, etc., were retrieved from laboratory measurements utilizing each model in order to be used for the simulations shown in this work.

### 4. Simulations and Model Comparison

With my family of semianalytical models I am going to choose scenarios and simulate a rocky surface for the intercomparison. The specific target is irrelevant since I am only comparing the models among them. Comparing with the results from other published theoretical frameworks of reflectance and emittance spectra will not be relevant either for the reason that at a given scenario one has to assume a mixture of materials in order to assemble a surface scenario. Those material proportions are not given for any of the published works in exoplanet science, to the author's best knowledge. I have selected LP 791-18d, a newly discovered Earth-sized planet orbiting the cool M6 dwarf LP 791-18 (I. J. M. Crossfield et al. 2019; M. S. Peterson et al. 2023) that is part of a coplanar system, which provides a unique opportunity to investigate a temperate exo-Earth in a system with a sub-Neptune that might have retained its volatile gas envelope. The gravitational interaction with the sub-Neptune prevents the complete circularization of LP 791-18d's orbit, resulting in continued tidal heating of the interior, which makes possible high volcanic activity. I am simulating the case that the volatile budget of LP 791-18d is exhausted. Near-future observations with JWST have been planned for this candidate in cycle 3 (B. Benneke et al. 2024) to distinguish at $4\sigma$ confidence between a bare-rock scenario and a $CO_2$-rich atmosphere that efficiently transports heat around the planets, even in the presence of Venus-like clouds. Observations are planned for five visits with F1500W filter and SUB256 subarray 3 $\mu$m wide bandpass centered at 15 $\mu$m, which covers a strong absorption feature from $CO_2$. For the model comparison and demonstration I am choosing four different scenarios with their compositions as follows, where the total sum should be 1 for 100%: (1) Ultramafic composed of 0.7 olivine, 0.2 pyroxene, and 0.1 chromite. (2) Feldspathic with plagioclase feldspar 0.75, Alkali feldspar 0.2, pyroxene 0.05, and olivine 0.05. (3) Metal-rich with iron 0.85, nickel 0.15, and cobalt 0.05. (4) Basaltic with plagioclase feldspar 0.4, pyroxene 0.3, olivine 0.2, and magnetite 0.1. For each one of those four scenarios I have used the laboratory measurements from the United States Geological Survey Spectral Library (USGS; R. Milliken 2020; R. E. Milliken et al. 2021; R. N. Clark et al. 2007), which typically are given for a limited set of incidence angles. I retrieve the optical and photometric parameters that are necessary for each model, as can be seen in Table 1, using the appropriate Hapke bidirectional model for each of the six approximations. To invert the problem and retrieve the wanted parameters, I applied the Bayesian algorithm that is described in

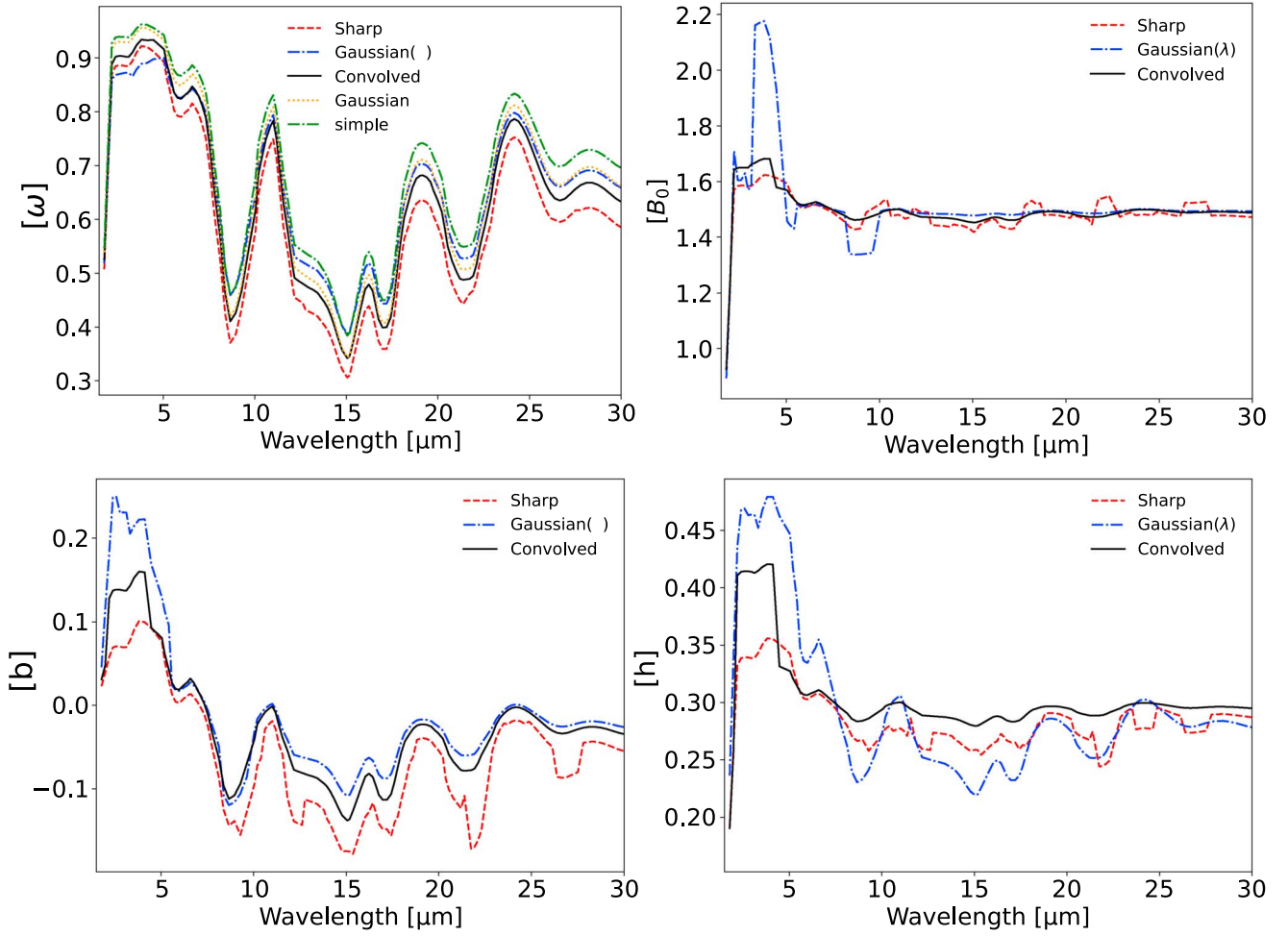

**Figure 1.** Retrieved optical properties of the ultramafic scenario. Parameter $\omega$ is the single-scattering albedo, $B_0$ is the amplitude of the surge term, $b$ is the scattering asymmetry factor, and $h$ is the angular width parameter that controls the spread of the opposition surge, as a sum of SHOE and CBOE, in phase-angle space.

Appendix B of L. Gkouvelis et al. (2024) and is based on minimization of $\chi^2$ by evaluating the Jacobians, $\partial F(x)/\partial x$, of the forward model, which in my case is the appropriate Hapke model. The total error is given by $\epsilon_{\text{total}}(\lambda) = \sqrt{\epsilon^2(\lambda) + \epsilon^2_{\text{noise}}(\lambda)}$, where $\epsilon(\lambda)$ represents the instrumental error for individual bands and $\epsilon_{\text{noise}}$ corresponds to the noise covariance matrix from the combined retrieval process of each spectral band set of measurements (C. D. Rodgers 2000). Comparison with minimization fitting packages of common astronomical software has shown small discrepancies. Figure 1 shows the retrieved optical properties of the ultramafic scenario, the $\omega$ for all models, and wavelength-dependent $B_0$, $h$, and $b$ opposition effect parameters for all the wavelength-dependent models. Single-scattering albedo retrieved across different models shows small discrepancies, remaining within the error estimation of $\approx 10\%$. $B_0$, $h$, and $b$ show overall good agreement and remain nearly constant for wavelengths $\gtrsim 5\,\mu$m. Similar behavior is observed for all materials considered in this work. For the models where the opposition effect parameters $B_0$, $h$, and $b$ are assumed constant across all wavelengths, I retrieve values of 1.5, $\approx 0.2$, and -0.1, respectively, for the ultramafic scenario. These values are close to those proposed by R. Hu et al. (2012), although their manuscript does not clearly explain how these values were determined. Thermal emission is the observation directly from the planet and can be detected as the thermal contrast of the host star as

$$\frac{F_p(\lambda)}{F_*(\lambda)} = \frac{f_p(\lambda)}{f_*(\lambda)}\frac{R_p^2}{R_*^2}\Phi(\alpha) + \left(\frac{R_p}{a}\right)^2 A_g \Phi(\alpha), \quad (32)$$

where $\Phi(\alpha)$ is the orbital phase angle of the planet. We can define $\alpha = 0$ in secondary eclipse when the observer looks directly on the dayside and $\alpha = 1$ at the primary transit, when the planet is in front of the host star. The second term is the contribution from scattered light, $\alpha$ is the semimajor axis of the star–planet system, and $A_g$ is the geometric albedo. I have expressed Equation (32) in the form of Equation (1). Nevertheless, from Equation (32) we can have a clear view that the planetary radiation flux in Equation (6) has to be evaluated at phase angle $\alpha = 0°$. However, for transiting systems the $\alpha = 0°$ is not observable owing to the planet contribution being obscured by the star (see, e.g., B. Placek et al. 2017). The secondary eclipse occurs when the planet's leading edge reaches the stellar limb, and it ends when the trailing edge fully exits, between phase angles $\alpha_{\text{eclipse}} \approx \pm \arcsin\left(\frac{R_* + R_p}{a}\right)$, with $a$ the semimajor axis. For LP 791-18d, utilizing the system parameters given in M. S. Peterson et al. (2023), the secondary eclipse lasts between $\pm 1°.81$.

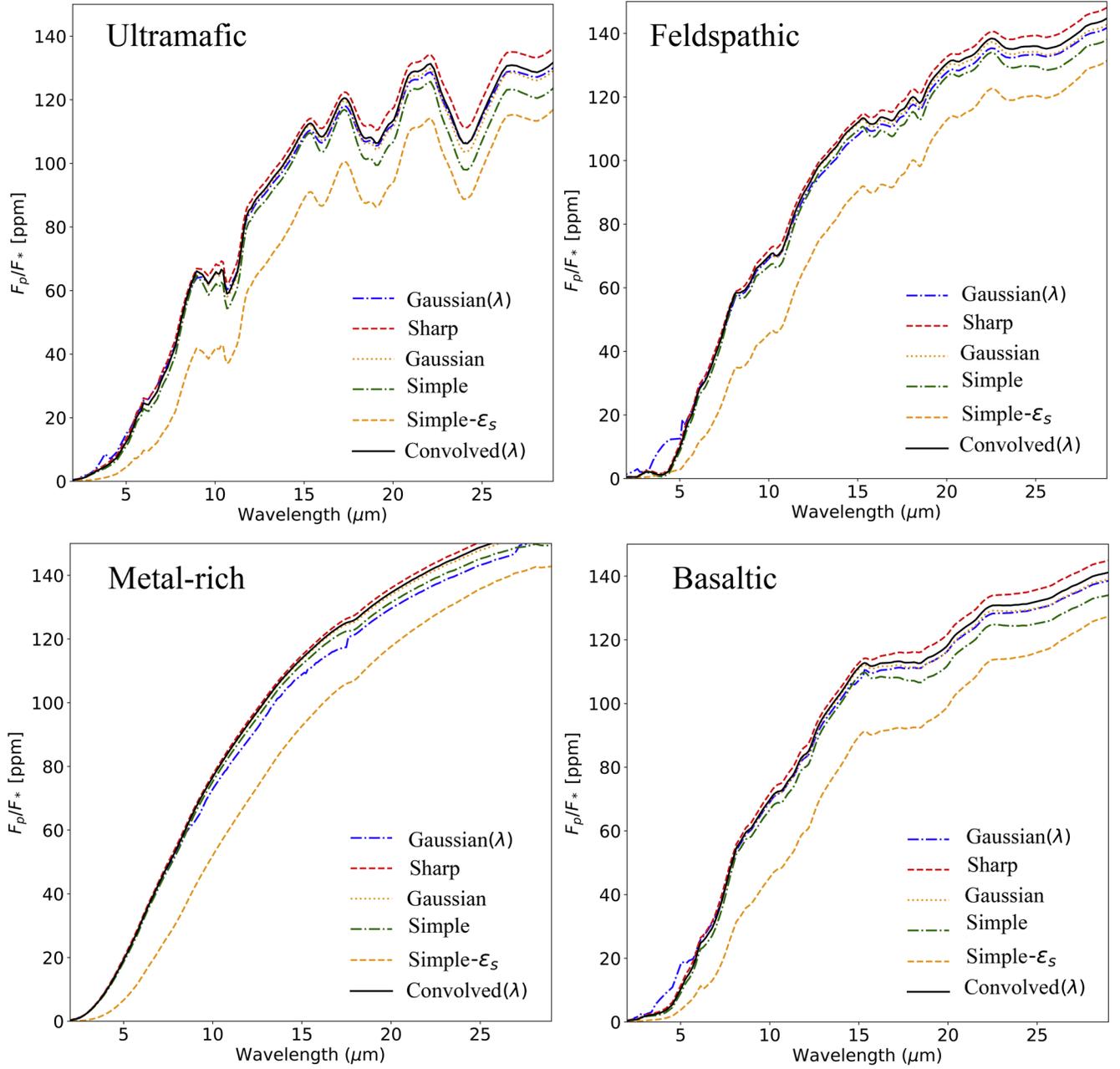

**Figure 2.** Comparison of model surface spectra of the case study rocky exoplanet LP 791-18d for four compositions: ultramafic, feldspathic, metal-rich, and basaltic.

In Figure 2 I present the synthetic spectra of secondary eclipse depth for the four scenarios per model. The deviations are generally small for wavelengths $\lesssim 15\,\mu m$, and for higher wavelengths they can reach $\pm 15\%$. The "simple-$\epsilon_s$" model deviates in comparison with all the other models and behaves like an outlier. Equation (20) is evaluated numerically, with convergence achieved at angular resolutions finer than approximately 7° for both $\mu$ and $\phi$.

In Figure 3, I plot the brightness temperature, $T_b(\lambda)$, as an indicator of the thermal emissivity of the planet (S. Seager & J. J. Lissauer 2010). This serves as a useful metric for direct comparisons with observations, but it is important to note that exoplanet observations capture the disk-averaged emission. The brightness temperature is obtained from the planet-to-star flux ratio by inverting the Planck function:

$$F_p = \pi B_\lambda(T_b)\left(\frac{R_p}{D}\right)^2. \tag{33}$$

Since the brightness temperature is influenced by the planet's total flux, the treatment of the opposition surge directly affects its retrieval. The opposition effect alters the surface reflectance, which modifies the partitioning between reflected and emitted radiation. As a result, models that neglect or approximate the opposition surge differently exhibit variations in inferred brightness temperature. This is particularly relevant for exoplanet thermal emission studies, where the brightness temperature is used as a proxy for surface temperature. The brightness temperature comparison shows generally small discrepancies across most models and wavelengths. However, some differences emerge in specific spectral regions. The wavelength-dependent Gaussian model presents instabilities at short wavelengths, where reflectivity is high for basaltic and metal-rich scenarios. The simple-$\epsilon$ model systematically underestimates brightness temperatures and deviates from all other models, particularly at longer wavelengths. This is consistent with the differences observed in the synthetic spectra in Figure 2, where neglecting the opposition surge modifies the planet's thermal emission. Additionally, the "Sharp" wavelength-dependent model exhibited computational instabilities, including nonphysical fluctuations and discontinuities, making it unreliable for practical applications. As a result, it is excluded from the plots. Finally, in Figure 4, I show the deviation of brightness temperature relative to its spectral mean. For the spectral averaging, I consider wavelengths $\gtrsim 7$ $\mu$m, as they are less affected by reflected light and provide a clearer view of thermal emission differences across models. This figure offers a valuable insight into intermodel discrepancies and highlights the influence of the opposition surge treatment on brightness temperature retrievals. A quantitative example for the ultramafic scenario is provided in Table 2, illustrating how different opposition models introduce systematic offsets in inferred surface properties.

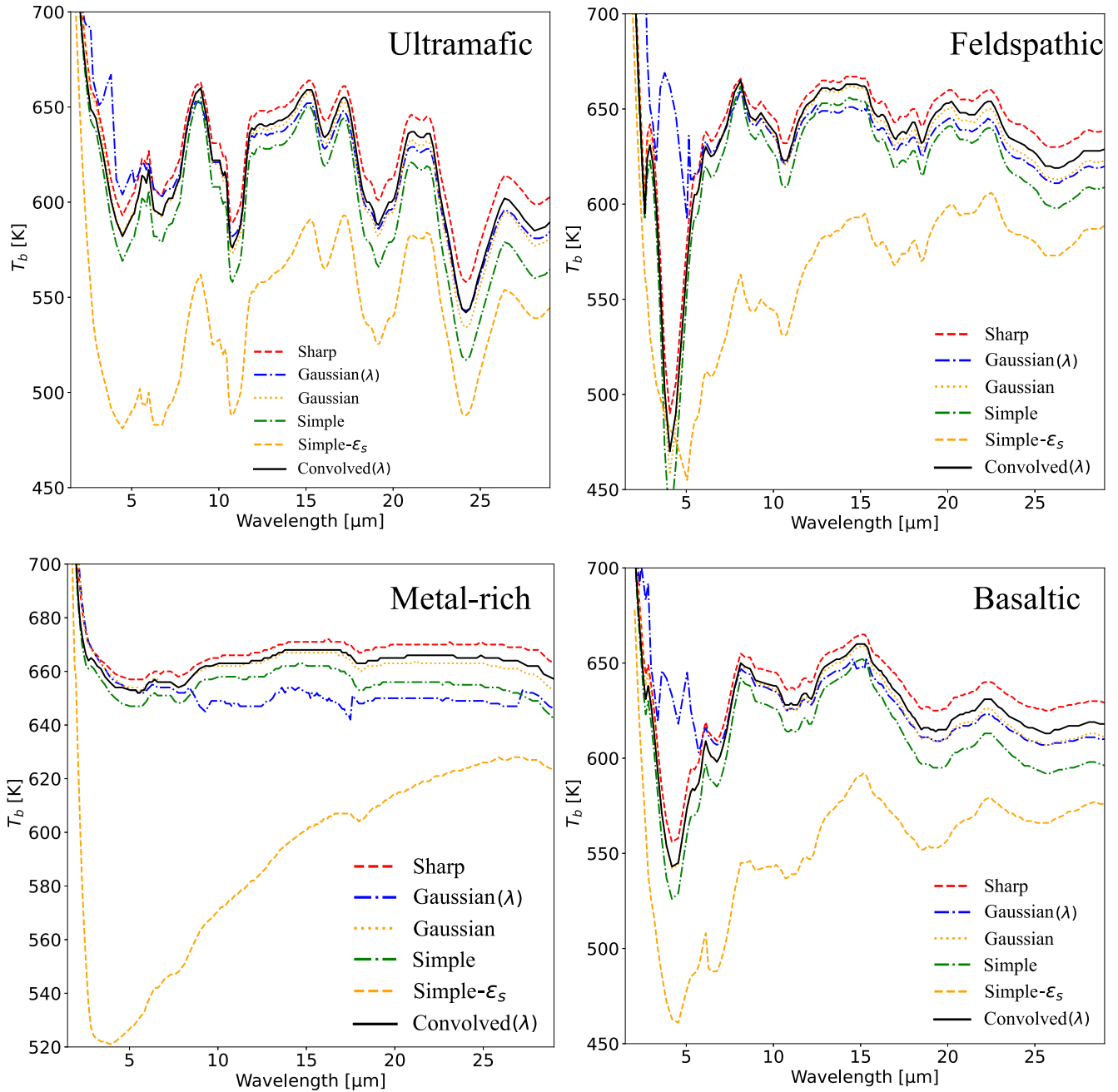

**Figure 3.** Brightness temperature for the four composition scenarios and each model. Brightness temperature can be directly compared with observations but also is an indicator of numerical behavior, when simulated. As an example, numerical instabilities can be seen in the Gaussian surge model for feldspathic and basaltic scenarios around 4–5 $\mu$m.

Phase curves of each scenario can be simulated, as is shown in Figures 5 and 6 for the ultramafic scenario in four selected bands, where the flux of $\pm 1$ $\mu$m is accounted for in each one of them.

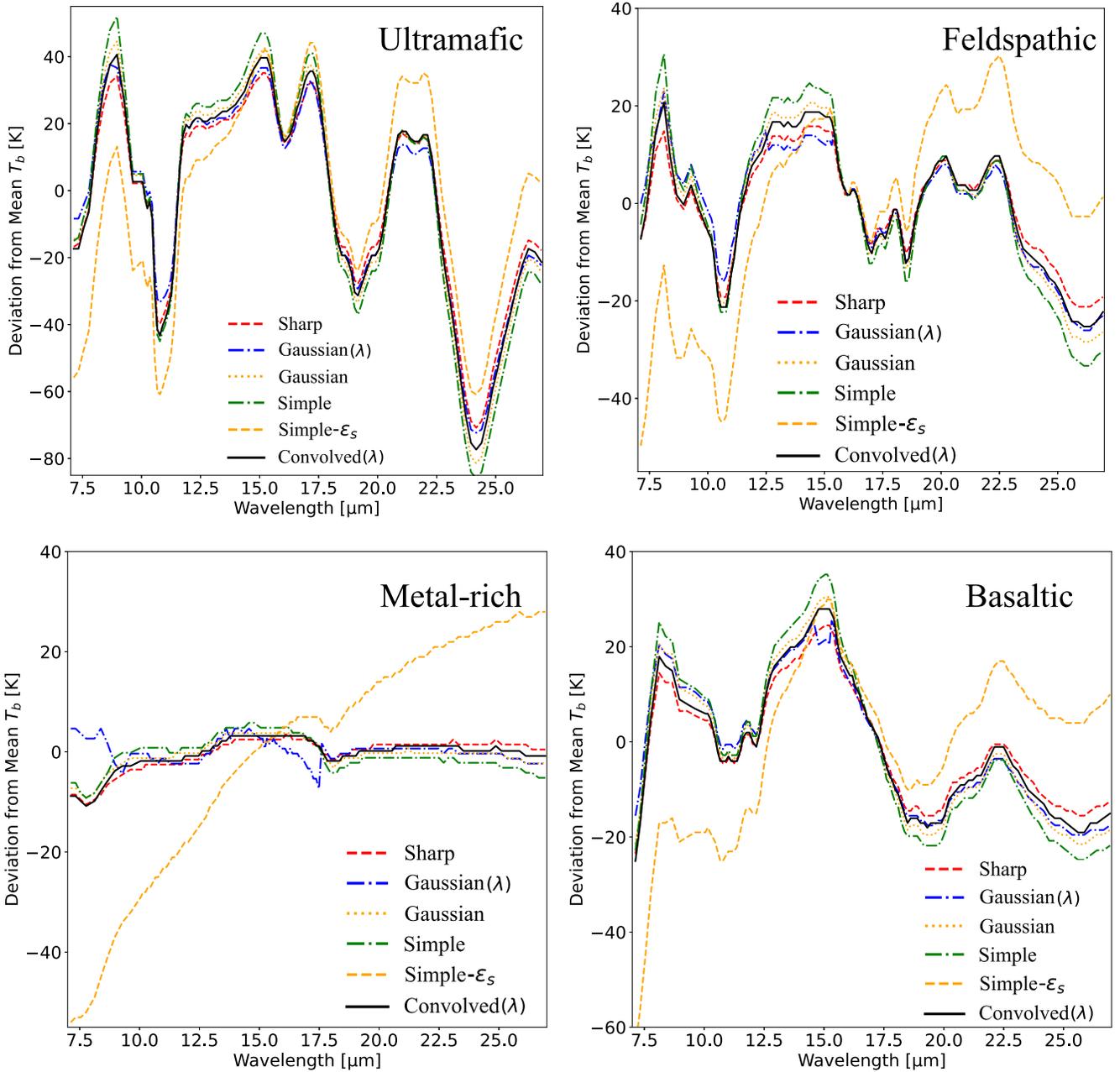

**Figure 4.** Deviation in brightness temperature. The deviation from the mean brightness temperature is shown for the spectral window 7–30 μm, which is an indicator of enhancement/suppression of spectral features from each model.

**Table 2**
Brightness Temperature for Ultramafic Model Comparison

| Name | Mean $T_b$ (K) | $\Delta T_b$ (K) | Deviation (K) |
| --- | --- | --- | --- |
| Sharp | 628 | 106 | 70 |
| Gaussian ($\lambda$) | 615 | 110 | 72 |
| Gaussian | 615 | 126 | 81 |
| Convolved ($\lambda$) | 619 | 108 | 69 |
| Simple | 603 | 138 | 86 |
| Simple-$\epsilon_s$ | 548 | 105 | 80 |

## 5. Conclusions and Discussion

I present a concise summary of reflectance and emittance spectroscopy applied to the surface of rocky exoplanets. My objective is to revisit the Hapke theory and optimize its formulation, with a focus on the airless rocky exoplanets. While observationally the opposition surge enhancement in brightness is an important factor for solar system objects across a wide spectral range, recent works in the field of exoplanets tend to neglect it (e.g., X. Lyu et al. 2024; M. Hammond et al. 2025). Given the current state of the observational precision in the exoplanet astronomy, this approximation is reasonable since the 10%–15% difference (for spectral bands $\gtrsim 10\,\mu$m) that I show in this work is within the uncertainty of the measurements. Nevertheless, it is important to have this difference in mind when we perform error assessment. Semiempirical works have shown that the opposition surge is the result of two natural phenomena, SHOE and CBOE, which I described in previous sections, and its contribution to the total brightness has a sharp or smooth peak around phase angle zero depending on material properties such as particle size, porosity, and crystallization. Its contribution can be mathematically described by Lorentzian- or Gaussian-style shape functions (E. Akkermans et al. 1988; D. G. Stankevich & Y. G. Shkuratov 2000; B. Hapke 2012), and its shape may vary with wavelength depending on the relative strength of SHOE and CBOE. The complete Hapke theory introduces these shapes ad hoc and requires the user to assume their mathematical form without prior knowledge, making accurate wavelength-dependent characterization challenging, particularly given the scarcity of comprehensive laboratory data. SHOE and CBOE are distinct phenomena but are not entirely independent. The physical characteristics of the material and surface—such as particle size, packing density, and optical properties—can interact in complex ways, influencing the effects. For example, larger particles tend to favor SHOE and reduce CBOE by limiting scattering paths. On the contrary, fine grains that are closely packed offer multiple scattering paths but limit SHOE owing to fewer shadows. Models that accurately describe opposition surges in solar system observations account for this interaction by adjusting the relative contribution of SHOE and CBOE according to particle characteristics (B. Hapke 1986; M. I. Mishchenko & J. M. Dlugach 1993; Y. Shkuratov & L. Starukhina 1999;, R. M. Nelson et al. 2000). Based on all the previous arguments, I have treated the opposition effect as a convolution of two mathematical functions. Convolution is often used to combine effects that are spatially or structurally dependent. The effects of SHOE and CBOE are both influenced by particle arrangements on the surface, and a convolution kernel can represent how one mechanism modulates the occurrence or effectiveness of the other. My approach can model both effects simultaneously and offer a unified, smooth transition across the phase-angle range and wavelength given that the optical properties of the material are known or retrieved from measurements. My objective is to develop a flexible mathematical tool that can fit the data properly across wavelengths, with the surge shape able to range from sharp to smooth or something in between, while remaining agnostic about the contribution or individual shape of each phenomenon. My target is not to separate the effects or show which of the two is more dominant or which mathematical function fits each one of them across the spectrum. I derived the convolved surge term in Equation (31), which modifies the bidirectional reflection formula (9) by replacing the ad hoc opposition enhancement term. I then demonstrated the performance of my approach by simulating the synthetic spectrum of the tidally locked exoplanet LP 791-18d as an airless rocky surface scenario in comparison with a family of models used in the literature. I assumed four surface scenarios composed of different materials that might be relevant for airless rocky exoplanets. I retrieved the necessary optical properties for each mixture by fitting the appropriate analytical model of the bidirectional reflectance. I highlighted the differences between various theoretical approximations by constructing a family of models that account for the opposition surge effect and others that do not. There is a systematic difference across all spectral ranges among the models. The relative difference varies from a few percent up to 30% depending on the wavelength band, as can be seen in Figure 5 for the 4 $\mu$m band. The wavelength dependence was tested in this type of theoretical framework applied in exoplanets for the first time, as well as intercomparisons of various models that have been used in the exoplanet community. Model "simple-$\epsilon_s$" is the one that oversimplifies the emissivity and shows the maximum deviation in eclipse depth and spectral shape, leading to an outlier behavior across all comparisons made in this work. Nevertheless, it is useful for fast calculations and Bayesian sampling in future analysis pipelines, as it is the least computationally expensive model, with a significant difference compared to all others. When obtaining thermal emission spectra of rocky exoplanets, without a priori knowledge of the existence of an atmosphere, a degeneracy may occur in the spectral analysis. For the atmospheric case, each spectral band is tracing different atmospheric layers following that $P_{\text{photosphere}} \approx g/\kappa$, with brightness temperature deviating far from equilibrium temperature. Similarly, I have shown in this work that under certain conditions an airless planet's surface can imitate the brightness temperature deviation of a hypothetical atmospheric layer.

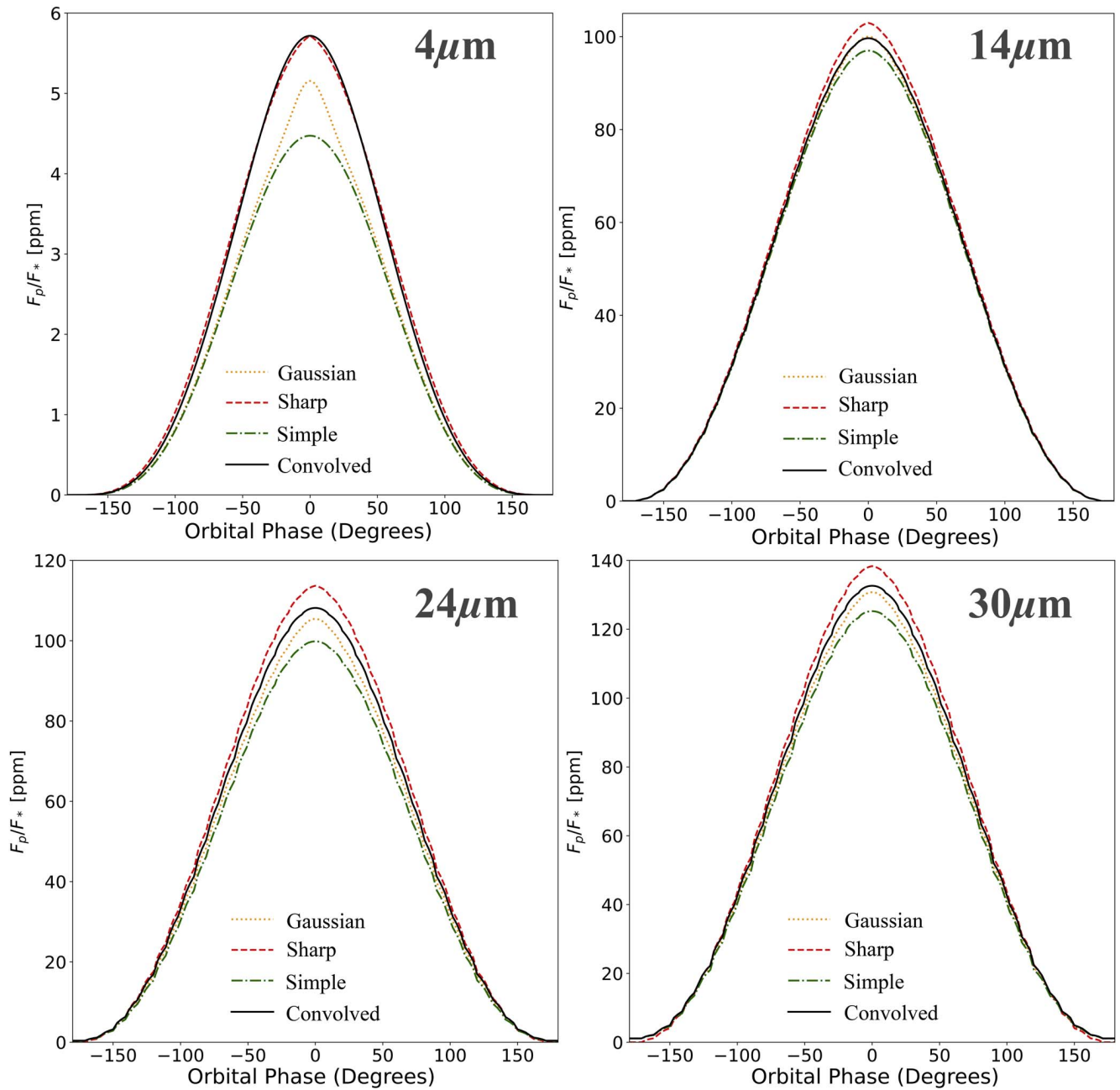

**Figure 5.** Phase curves of the ultramafic scenario for selected spectroscopic bands. Four basic model phase curves are presented, where the black solid curve is the new approach for opposition surge that has the ability to adapt from a sharp to a smooth profile, according to ultramafic optical properties.

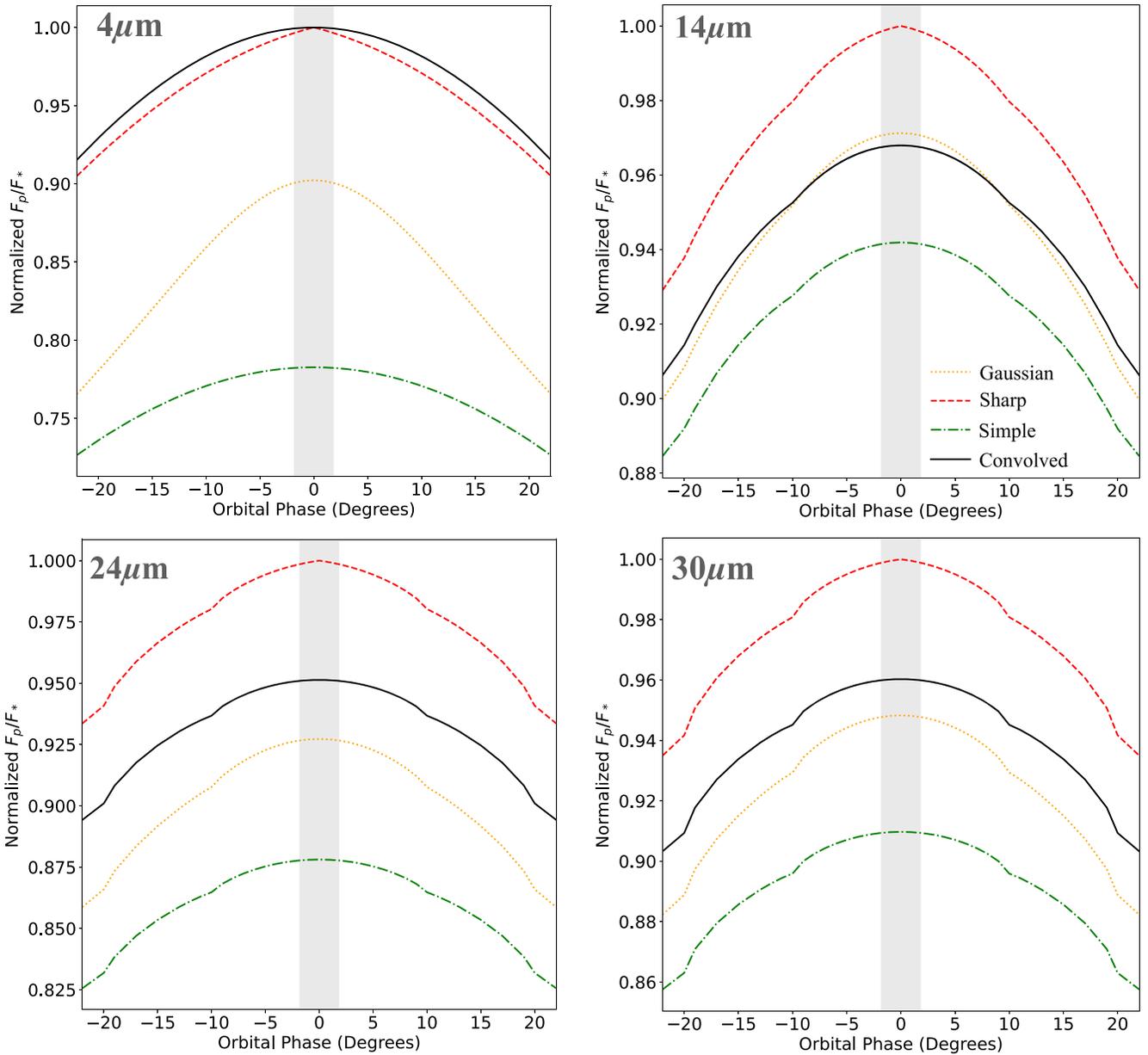

**Figure 6.** Same as Figure 5, but zoomed in ±20° around secondary eclipse and normalized by the highest value to emphasize the model differences. With the gray stripe LP 791-18d's secondary eclipse total duration of ≈3°.62 is indicated.

*5.1. Mitigating Refinement Challenges and Envisioning Future Applications*

As a first approach, from this work I conclude that for exoplanet science omitting the opposition surge effect will introduce uncertainties that are not important when compared to the uncertainty in the current observational data. Nevertheless, the opposition surge is a phenomenon that should be of great importance in the interpretation of future observations. For example, the Habitable Worlds Observatory (HWO) will search for biosignatures in Earth-size exoplanets of nearby stars with high-resolution spectroscopy in near-ultraviolet, visible, and near-infrared wavelengths with high reflectivity (M. Damiano & R. Hu 2023; R. Morgan et al. 2025).

I have to note at this point that for each of the material samples mostly a limited set of angles was found in the laboratory measurements for the bidirectional reflectance. These data sets might cause uncertainties and degeneracies in the retrieved optical and photometric properties of the material. Especially for materials for which only a single incident angle is measured (e.g., $i \geqslant 30°$), only the single-scattering albedo can be retrieved, and no parameters relevant to the opposition surge can be constrained. With this work, I aim to stimulate a creative interplay between exoplanet scientists and experimental physicists who perform such measurements, in order to envision and design future experiments and enrich spectral databases in ways that optimize the scientific return of next-generation telescopes focusing on rocky planets. This type of theoretical framework is gaining popularity within the exoplanet science community, as part of efforts to distinguish between atmospheric signatures and the airless "bare-rock" degeneracy. It is the author's educated guess that such models will be widely used in the near future for the characterization of rocky worlds. Current efforts primarily focus on characterizing giant planets and rely on Bayesian inference techniques. These frameworks combine atmospheric forward models with atomic and molecular measurement databases (e.g., K. L. Chubb et al. 2024) to sample parameter space and generate probability distributions for atmospheric composition and thermal structure. Similarly, in the upcoming years—when technology and data permit—for surface characterization, the same Bayesian statistical machinery should be applied. This would involve utilizing the theoretical frameworks described in this work, with the resulting probability distributions yielding estimates of surface composition, grain size, porosity, and MRBB. Additionally, while my analysis focuses on SHOE and CBOE, I acknowledge that recent work by D. J. Shiltz & C. M. Bachmann (2023) highlights the importance of MRBB effects. My work utilizes laboratory measurements of powders without the MRBB effect. Nevertheless, since the MRBB acts across the entire hemisphere, including at small phase angles, future models should incorporate a separate treatment of this effect to refine opposition surge predictions for exoplanetary surfaces.

## Acknowledgments

I would like to thank the two anonymous reviewers for the constructive feedback. I am also grateful to my family—Laura, Sophie, and Aris—for their patience and support throughout the course of this project.

## ORCID iDs

Leonardos Gkouvelis  https://orcid.org/0000-0002-1397-8169